\documentstyle[12pt,epsf]{article}

\newcommand{\alt}{\mathrel{\raisebox{-.6ex}{$\stackrel{\textstyle<}{\sim}$}}}
\newcommand{\agt}{\mathrel{\raisebox{-.6ex}{$\stackrel{\textstyle>}{\sim}$}}}
\def\overlay#1#2{\ifmmode \setbox 0=\hbox {$#1$}\setbox 1=\hbox to\wd 0{\hss
$#2$\hss }\else \setbox 0=\hbox {#1}\setbox 1=\hbox to\wd 0{\hss #2\hss }\fi
#1\hskip -\wd 0\box 1}
\def\case#1/#2{{\textstyle{#1\over#2}}}
\catcode`@=11
\newcount\@tempcntc
\def\@citex[#1]#2{\if@filesw\immediate\write\@auxout{\string\citation{#2}}\fi
  \@tempcnta\z@\@tempcntb\m@ne\def\@citea{}\@cite{\@for\@citeb:=#2\do
    {\@ifundefined
       {b@\@citeb}{\@citeo\@tempcntb\m@ne\@citea\def\@citea{,}{\bf ?}\@warning
       {Citation `\@citeb' on page \thepage \space undefined}}%
    {\setbox\z@\hbox{\global\@tempcntc0\csname b@\@citeb\endcsname\relax}%
     \ifnum\@tempcntc=\z@ \@citeo\@tempcntb\m@ne
       \@citea\def\@citea{,}\hbox{\csname b@\@citeb\endcsname}%
     \else
      \advance\@tempcntb\@ne
      \ifnum\@tempcntb=\@tempcntc
      \else\advance\@tempcntb\m@ne\@citeo
      \@tempcnta\@tempcntc\@tempcntb\@tempcntc\fi\fi}}\@citeo}{#1}}
\def\@citeo{\ifnum\@tempcnta>\@tempcntb\else\@citea\def\@citea{,}%
  \ifnum\@tempcnta=\@tempcntb\the\@tempcnta\else
   {\advance\@tempcnta\@ne\ifnum\@tempcnta=\@tempcntb \else \def\@citea{--}\fi
    \advance\@tempcnta\m@ne\the\@tempcnta\@citea\the\@tempcntb}\fi\fi}
\catcode`@=12

\renewenvironment{thebibliography}[1]
 {\begin{list}{\arabic{enumi}.}
    {\usecounter{enumi} \setlength{\parsep}{0pt}
     \setlength{\itemsep}{3pt} \settowidth{\labelwidth}{#1.}
     \sloppy
    }}{\end{list}}

\def\mm{\mu^+\mu^-}
\def\ee{e^+e^-}
\def\rta{\rightarrow}
\def\tanb{\tan\beta}

\begin{document}
\baselineskip13.5pt

\font\fortssbx=cmssbx10 scaled \magstep2
\hbox to \hsize{
\raise.1in\hbox{\fortssbx University of Wisconsin - Madison}
\hfill\vbox{\hbox{\bf MADPH-95-873}
            \hbox{March 1995}}}

\begin{center}
{\large\bf
Physics Goals of a{\boldmath$\mu^+\mu^-$} Collider\footnotemark }\\[.1in]
\small
V.~Barger$^a$, M.S.~Berger$^b$, K.~Fujii$^c$, J.F.~Gunion$^d$,
T.~Han$^d$, C.~Heusch$^e$, W.~Hong$^f$,
S.K.~Oh$^g$, Z.~Parsa$^h$, S.~Rajpoot$^i$, R.~Thun$^j$, and
B.~Willis$^k$\\[.1in]
\small\it
$^a$Physics Department, University of Wisconsin, Madison, WI 53706,
USA\\
$^b$Physics Department, Indiana University, Bloomington, IN 47405,
USA\\
$^c$Physics Division, KEK,  1-1 Oho, Tsukuba, Ibaraki, Japan\\
$^d$Physics Department, University of California,  Davis, CA 95616, USA\\
$^e$Physics Department, University of California,  Santa Cruz, CA 95064, USA\\
$^f$Physics Department, University of California,  Los Angeles, CA 90024, USA\\
$^g$Physics Department, Kon-kuk University, Seoul 133-701, South Korea\\
\mbox{$^h$Physics Department, Brookhaven National Laboratory, Upton, NY 11973,
USA}\\
\mbox{$^i$Physics Department, California State University, Long Beach, CA
90840,
USA}\\
$^j$Physics Department, University of Michigan, Ann Arbor, MI 48109, USA\\
$^k$Physics Department, Columbia University, New York, NY 10027,
USA
\end{center}

\footnotetext{
Report of Physics Goals Working Group presented by V. Barger at the {\it 2nd
Workshop on Physics Potential and Development of $\mu^+\mu^-$ Colliders},
Sausalito, California, Nov.~1994. Working group leaders V.~Barger and
J.F.~Gunion.}

\vspace*{-.2in}

\begin{abstract}
\vspace*{-.1in}

\noindent
This working group report focuses on the physics potential of $\mu^+\mu^-$
colliders beyond what can be accomplished at linear $e^+e^-$ colliders and the
LHC. Particularly interesting possibilities include (i)~$s$-channel resonance
production to discover and study heavy Higgs bosons with $ZZ$ and $WW$
couplings that are suppressed or absent at tree-level
(such as the $H$ and $A$ Higgs bosons of supersymmetric models),
(ii)~study of the strongly interacting electroweak sector, where higher
energies give larger signals, and (iii)~measurements of the masses and
properties of heavy supersymmetric particles.

\end{abstract}

\smallskip

\begin{center}
{\large\bf\uppercase{I. Introduction}}
\end{center}

\vspace{-.1in}

The physics working group focused on the new physics potential of a $\mm$
collider\cite{mupmum1}.
Since the time available for analysis was limited, this report is
largely directed to two areas of major current interest in particle physics:

\begin{itemize}
\addtolength{\itemsep}{-.075in}

\item finding Higgs bosons or detecting
strong $WW$ scattering and thereby understanding the
origin of electroweak symmetry breaking (EWSB);

\item finding and studying supersymmetric (SUSY) particles.

\end{itemize}

\noindent
For the same energy and integrated luminosity, it should be possible to do
anything at a $\mm$ collider that can be done at
an $\ee$ collider. Moreover, a $\mm$ collider opens up two particularly
interesting possibilities for improving the physics reach
over that of $\ee$ colliders:

\begin{itemize}

\item $s$-channel Higgs production;
\item higher center-of-mass energy with reduced backgrounds.

\end{itemize}

\noindent Both possibilities are simply a result of the large
muon mass as compared to the electron mass. Direct $s$-channel Higgs
production is greatly enhanced at a $\mm$ collider because
the coupling of
any Higgs boson to the incoming $\mm$ is proportional to $m_{\mu}$, and
is therefore much larger than in $\ee$ collisions.  Higher
energy is possible at a $\mm$ collider due to the ability
to recirculate the muons through a linear acceleration array without
an overwhelming radiative energy loss. Current estimates are that
an $\ee$ collider with energy above 1.5 TeV will be very difficult
to construct (with adequate luminosity), whereas a 4 TeV $\mm$ collider
would appear to be well within the range of possibility \cite{const}.
Meanwhile, backgrounds from processes arising from photon radiation
from one or both beams will be suppressed by the higher muon mass.
As we shall discuss, higher energy could be crucial in at least two ways:

\begin{itemize}

\item improved signals for strong $WW$ scattering --- since higher energies are
achievable the signal level is increased, while backgrounds are reduced
due to there being less photon radiation;

\item the kinematical reach for pair production of SUSY particles
is extended to a possibly crucial higher mass range.

\end{itemize}

Very briefly, what new physics can an $\ee$ collider\cite{epem} do?
First, neutral Higgs
bosons can be discovered that are coupled to the $Z$-boson. The Standard
Model Higgs boson or the lightest Higgs boson of the Minimal Supersymmetric
Standard Model (MSSM) can be discovered (in $Z^*\rta Z h$ production)
if $m_h \alt 0.7 \sqrt s$. Since there is a
theoretical upper bound\cite{higgs1,higgs2}
on the lightest Higgs in the MSSM of $m_h\alt130$ to
150~GeV, an $\ee$ machine with CM energy $\sqrt s=300$~GeV can exclude or
confirm supersymmetric theories based on grand unified theories (GUTs).
However, if heavy, the other neutral SUSY Higgs bosons $H,A$ can only be
discovered via $Z^*\rta HA$ production for $m_H\sim m_A < \sqrt s/2$;
$Z^*\to ZH$ ($ZA$) production is not useful since
the $H$ ($A$) coupling to $ZZ$ is suppressed (absent)
at tree-level.

Discovery of SUSY sparticles is also
possible at an $\ee$ collider for sparticle masses $m<\sqrt
s/2$. While the energy reach of an $\ee$ collider will probably be
adequate for pair producing the lightest chargino $\chi_1^\pm$,
the neutralino combination $\chi_1^0\chi_2^0$ and possibly the
selectron, smuon and stau $\tilde e,\tilde\mu,\tilde\tau$,
and the lighter stop eigenstate, $\tilde t_1$,
it could be inadequate for the heavier chargino and neutralinos and other
squarks.

The figure of merit in physics searches at an $\ee$ or $\mm$ collider is the
QED point cross section for $\ee\to\mm$, which has the value
\begin{equation}
\sigma_{QED} (\sqrt s) = {100{\rm\ fb}\over s\rm\ (TeV^2)}
\left(\alpha(s)\over\alpha(M_Z^2)\right)^2 \;.
\end{equation}
Henceforth we neglect the factor $\left(\alpha(s)/\alpha(M_Z^2)\right)^2$,
which varies slowly with $s$. As a rule of thumb, the integrated luminosity
needed for the study of new physics signals is
\begin{equation}
\left( \int {\cal L} dt \right) \sigma_{QED} \agt 1000\rm\ events \;.
\end{equation}
Thus $\mm$ machine designs should be able to deliver an
integrated luminosity of
\begin{equation}
\int {\cal L} dt \agt 10 \cdot s \rm\ (fb)^{-1} \;.
\end{equation}
If this is to be accumulated in one year's running, the luminosity requirement
is
\begin{equation}
{\cal L} \agt 10^{33} \cdot s\rm\ (cm)^{-2} \, (sec)^{-1} \;.
\end{equation}
Two possible $\mm$ machines were considered at this meeting:

\begin{itemize}

\item Low $\sqrt s \simeq 400$ GeV, requiring
\begin{equation}
\int {\cal L}dt \agt 1\rm\ (fb)^{-1}, \quad
{\cal L} \agt 10^{32}\rm\ (cm)^{-2}\, (sec)^{-1}
\end{equation}

\item High $\sqrt s \agt 4$ TeV, requiring
\begin{equation}
\int {\cal L}dt \agt 100\rm \ (fb)^{-1}, \quad
{\cal L} \agt 10^{34}\rm\ (cm)^{-2}\, (sec)^{-1}
\end{equation}

\end{itemize}

\noindent
Fortunately these luminosity requirements may be achievable with the
designs under consideration~\cite{const}. Indeed,
the luminosity of a $\sqrt s \sim
400$ GeV machine could be as large as ${\cal L}\agt
10^{33}\rm\ (cm)^{-2}\, (sec)^{-1}$.

\bigskip
\begin{center}
{\large\bf\uppercase{II. First Muon Collider} (FMC)}
\end{center}

Consider a $\mm$ machine with CM energy $\sqrt s$
in the range 250 to 500 GeV.
As noted above, the yearly integrated luminosity
for this machine would be at least $L\geq 1~\rm(fb)^{-1}$ and possibly
as much as $L\geq 10-20~\rm(fb)^{-1}$.
The most interesting physics at
such a $\mm$ collider that goes beyond that accessible
at an $\ee$ collider of similar energy is
the possibility of $s$-channel heavy Higgs production, as
illustrated in Fig.~1.

\begin{center}
\epsfxsize=2.25in\hspace{0in}\epsffile{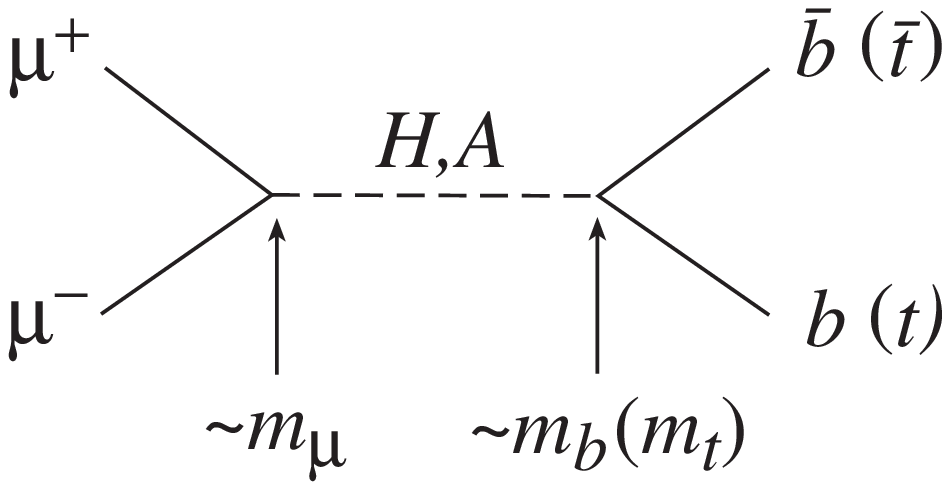}

\medskip
{\footnotesize\sf Fig.~1: $s$-channel diagrams for production of $H,A$ MSSM
Higgs
bosons.}
\end{center}

\noindent
To discover a narrow resonance, one would ideally like to have a broad
$\sqrt s$ spectrum.  (Alternatively,
a scan with limited luminosity at discretely spaced
energies could be employed; our results
for the case of a broad spectrum are easily altered to this latter
procedure.) Once a Higgs boson is discovered, it would be best
to change to as narrow as possible a spectrum and then sit on the peak to study
the resonance properties.

The $s$-channel Higgs resonance cross section is
\begin{equation}
\sigma_h = {4\pi \Gamma(h\to\mu\mu) \, \Gamma(h\to X)\over
\left(s-m_h^2\right)^2 + m_h^2 \Gamma_h^2} \;,
\end{equation}
where $X$ denotes a final state and $\Gamma_h$ is the total width. The
sharpness of the resonance peak is determined by $\Gamma_h$. For
a broad energy spectrum, the relevant
energy resolution is that determined by the detector.  The widths
of the Higgs
bosons under consideration will be seen to be quite small, such that
it is reasonable to consider
\begin{equation}
\Delta E({\rm resolution}) \geq \Gamma(\rm Higgs) \;.
\end{equation}
Then the integrated signal over the narrow resonance is
\begin{equation}
S_h = \int \sigma_h d\sqrt s = {\pi\over2} \Gamma_h \sigma_h^{\rm peak} \;,
\end{equation}
where the peak cross section is
\begin{equation}
\sigma_h^{\rm peak} = {4\pi\over m_h^2} {\rm BF}(h\to\mu\mu)
{\rm BF}(h\to X)\;,
\end{equation}
leading to
\begin{equation}
S_h = {2\pi^2\over m_h^2} \Gamma(h\to\mu\mu) \,{\rm BF}(h\to X) \;.
\end{equation}
With the integrated luminosity $L = \int {\cal L}dt$ spread over  an energy
band $\Delta E$ in the search mode, the event  rate for the Higgs signal is
\begin{equation}
N_h = S_h {L\over \Delta E} \;.
\end{equation}
In the above, we have used the general notation $h$ for the Higgs boson.

In applying the above general formulae to the Minimal Supersymmetric Model
(MSSM) Higgs bosons we note the following important
facts. The couplings to fermions and vector bosons
depend on the SUSY parameter
$\tan\beta=v_2/v_1$ and on the mixing angle $\alpha$ between the neutral Higgs
states ($\alpha$ is determined by the Higgs masses, $\tan\beta$, the top and
the stop masses). SUSY GUT models predict large $m_A$ and
$\alpha\approx\beta-\pi/2$.  In this case,
the coupling factors of the Higgs bosons are approximately\cite{gh}
\begin{equation}
\begin{array}{lccc}
& \mm, b\bar b& t\bar t& ZZ,W^+W^-\\
h & 1 & -1 & 1\\
H & \tan\beta & -1/\tan\beta & 0 \\
A & -i\gamma_5\tan\beta & -i\gamma_5/\tan\beta & 0
\end{array}
\end{equation}
times the Standard-Model factor of $g m/(2m_W)$  in the case of fermions (where
$m$ is the relevant fermion mass), or $gm_W,gm_Z/\cos\theta_W$
in the case of the $W,Z$.
The broad spectrum inclusive signal rate $S_h$ is proportional to $\Gamma(h\rta
\mm)$ and since
the coupling of $h=H,A$ to the $\mm$ channel is proportional to $\tan\beta$,
larger $\tan\beta$ values give larger production.

To obtain the rate
in a given final state mode $X$ we multiply the inclusive rate
by BF$(h\rta X)$. Here, we consider only the
$b\bar b$ and  $t\bar t$ decay modes for $h=H,A$, although
the relatively background free $H\rta hh\rta b\bar
b b\bar b$ and $A\rta Zh\rta Zb\bar b$ modes might also be useful for
discovery. Figure 2
shows the dominant branching fractions to $b\bar b$ and $t\bar t$ of
Higgs bosons of mass $m_A=400{\rm\ GeV}\approx m_H$ versus $\tan\beta$, taking
$m_t=170$~GeV. The $b\bar b$ decay mode is dominant for $\tan\beta>5$,
which is the region where observable signal rates are obtained. From the
figure we see that BF$(h\rta b\bar b)$ grows rapidly with $\tanb$ for
$\tanb\alt 5$, while BF$(h\rta t\bar t)$ falls slowly. For low to moderate
$\tanb$ values, the event rates behave as
\begin{eqnarray}
N(\mm\to H,A\to b\bar b) &\propto& m_\mu^2 m_b^2\left[
(\tan\beta)^6\mbox{ to }(\tan\beta)^4 \right]\\
N(\mm\to H,A\to t\bar t) &\propto&  m_\mu^2 m_t^2\left[
(\tanb)^2\mbox{ to }(\tanb) \right]\,.
\end{eqnarray}
It is this growth with $\tanb$ that makes $H,A$ discovery
possible for relatively modest values of $\tanb$ larger than 1.
At high $\tanb$ the $b\bar b$ branching ratio asymptotes
to a constant value while the $t\bar t$ branching ratio falls as $1/(\tanb)^4$,
so that
\begin{eqnarray}
N(\mm\to H,A\to b\bar b) &\propto& m_\mu^2 m_b^2(\tan\beta)^2\\
N(\mm\to H,A\to t\bar t) &\propto&  m_\mu^2 m_t^2(\tanb)^{-2}\,.
\end{eqnarray}
Consequently, the $t\bar t$ channel will not be useful at large $\tanb$,
whereas the $b\bar b$ channel continues to provide a
large event rate.

\begin{center}
\epsfxsize=3.5in\hspace{0in}\epsffile{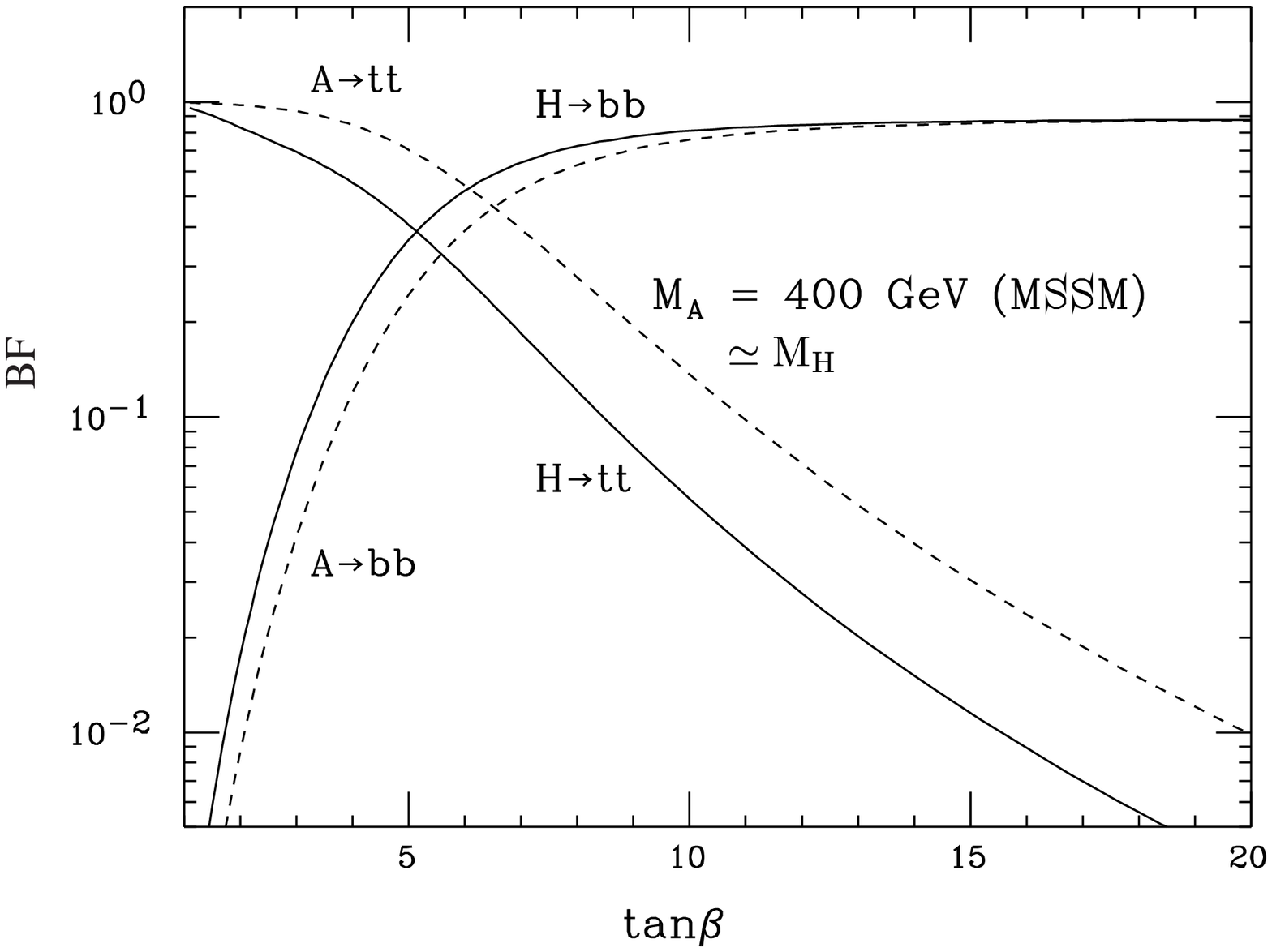}

\medskip
\parbox{5.5in}{\footnotesize\sf Fig.~2: Dependence of the branching fractions
for the heavy  supersymmetric Higgs bosons on $\tan \beta$ (from
Ref.~\cite{trans}).}
\end{center}

The calculated Higgs boson
widths are shown in Fig.~3 versus $m_h$ for $\tan\beta=5$.
As promised, the $H$ and
$A$ are typically narrow resonances ($\Gamma_{H,A}\sim 0.1$ to 2~GeV),
and our approximation of $\Delta E \geq \Gamma_{H,A}$ will generally
be valid; see below.

\begin{center}
\epsfxsize=3.5in\hspace{0in}\epsffile{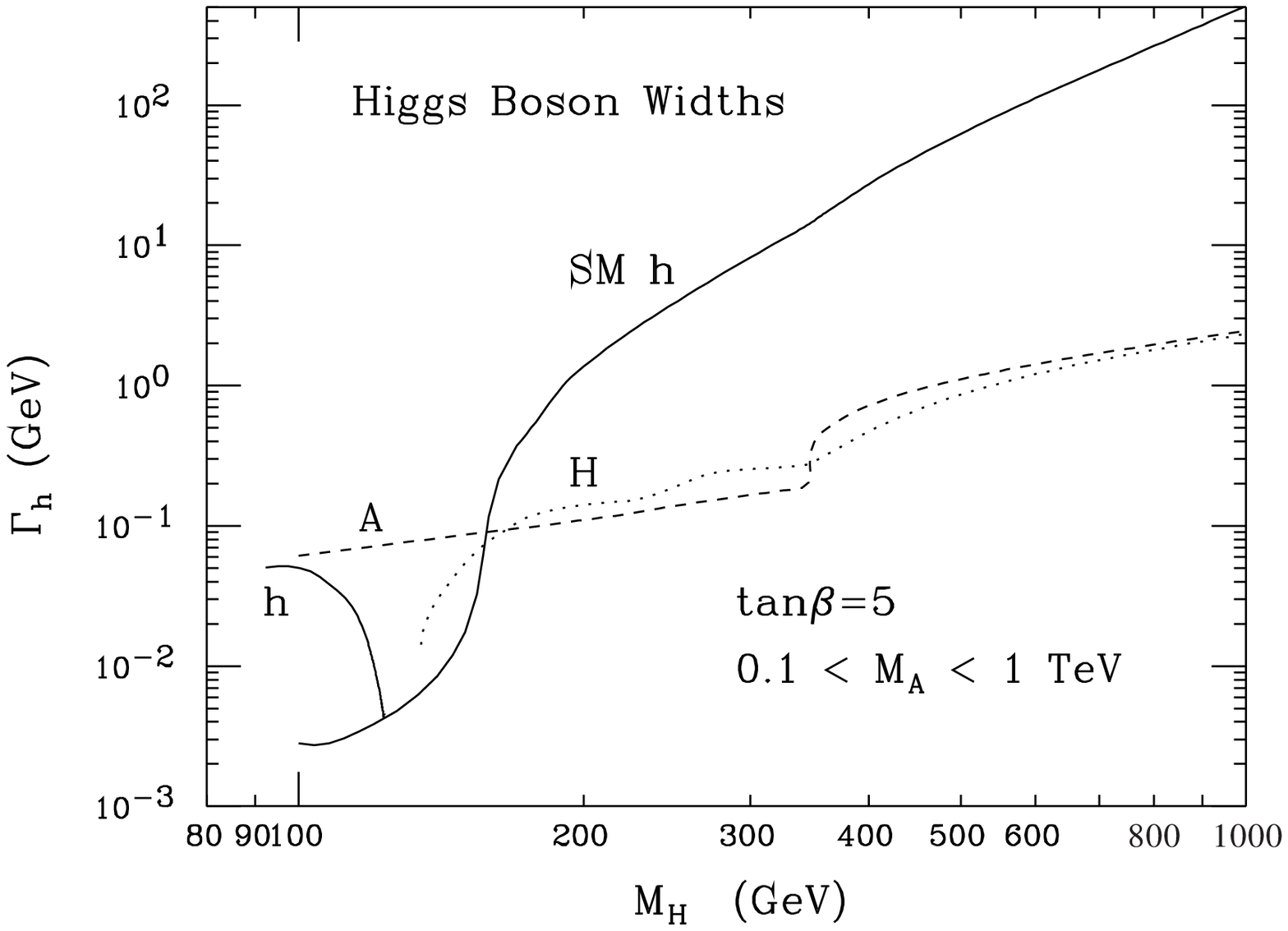}

\medskip
\parbox{5.5in}{\footnotesize\sf Fig.~3: The Standard Model Higgs boson and the
supersymmetric Higgs
boson widths (from Ref.~\cite{trans}).}
\end{center}

The irreducible backgrounds to the Higgs signals are
\begin{equation}
\mm \to \gamma^*, Z^* \to b\bar b, t\bar t \;.
\end{equation}
The final-state energy resolution can be estimated by taking the heavy quark
energy resolution to be that of the hadron energy resolution in NLC design
studies~\cite{epem}
\begin{equation}
{\Delta E_Q\over E_Q} = {50\%\over \sqrt{E_Q}} + 2\% \;.
\end{equation}
Then for example at $m_h=400$~GeV the final state mass resolution is
\begin{equation}
\Delta m(b\bar b) \simeq 2\% m(b\bar b)\;.
\end{equation}
The background cross section is integrated over the bin $\sqrt
s=m_h\pm{1\over2}\Delta m(b\bar b)$
\begin{equation}
B = \int \sigma_B \, d\sqrt s \;.
\end{equation}
The light-quark backgrounds can be rejected using $b$-tagging. As a single
$b$-tag efficiency, we assume $\epsilon_b\simeq0.5$ and neglect mistags.

The $H$ signal (for $\tan\beta=5$) and the backgrounds integrated over the
resolution are shown in Fig.~4
versus $m_H$. As a concrete example let us
assume the optimistic integrated luminosity of $L=20$~(fb)$^{-1}$
spread over an energy band
$\Delta E=50$~GeV, giving $L/\Delta E=0.4\rm~(fb)^{-1}/GeV$. The signal in the
$b\bar b$ channel for $m_H=400$~GeV with $\tan\beta=5$ is
\begin{equation}
N({\rm signal}) = 250\rm\ events   \label{Nsignal}
\end{equation}
while the $b\bar b$ background is
\begin{equation}
B({b\bar b \; {\rm background}}) = 2000\rm\ events \;,
\end{equation}
where both numbers include the single $b$-tag efficiency $\epsilon_b$.
The significance of the signal is
\begin{equation}
n_{SD}  \simeq 5.6 \;.  \label{nSD}
\end{equation}
Thus, discovery is possible here!

\begin{center}
\epsfxsize=3.5in\hspace{0in}\epsffile{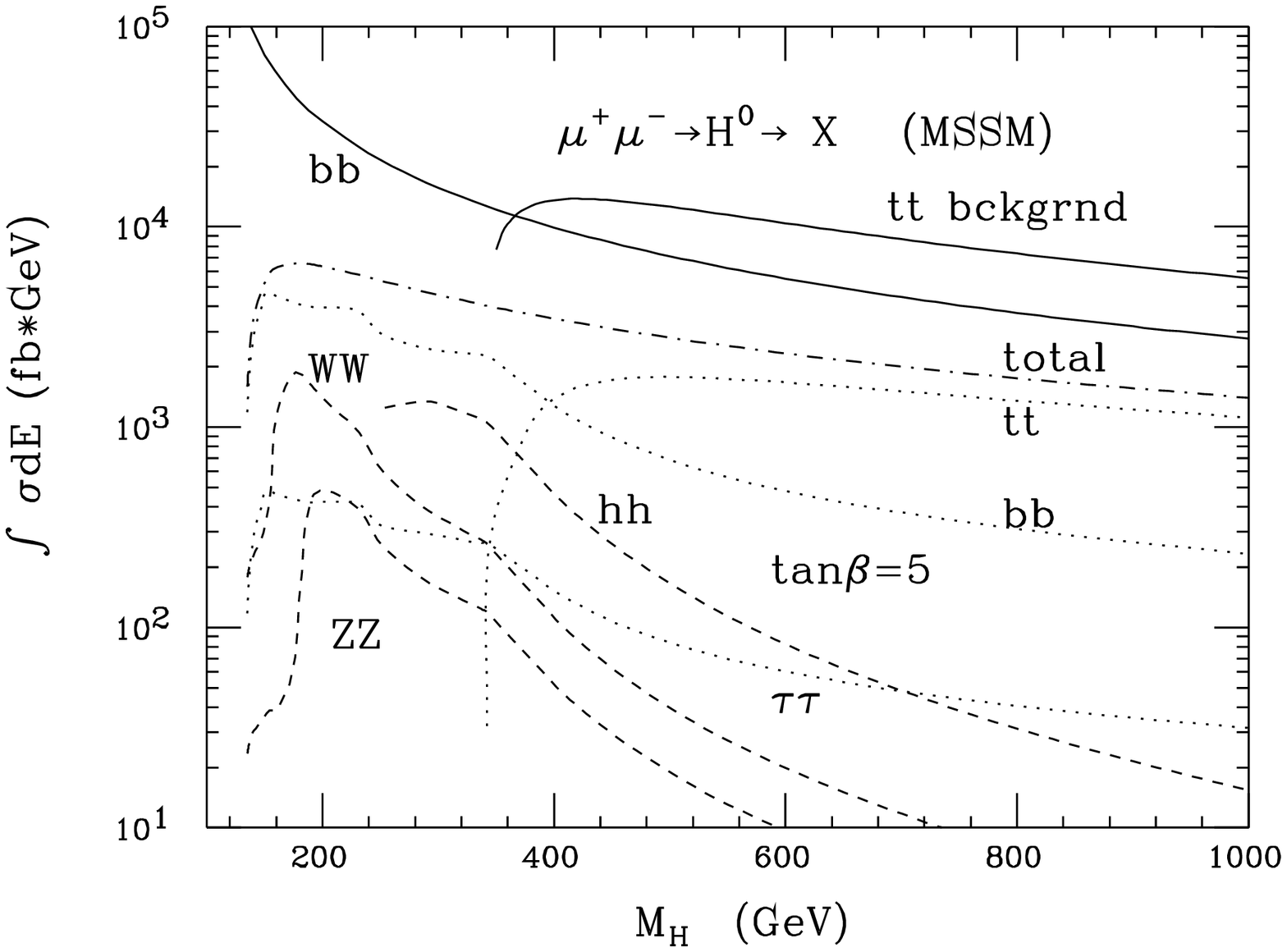}

\smallskip
{\footnotesize\sf Fig.~4: Typical signals and backgrounds for $\mu^+\mu^-\to H
\to X$ in
the MSSM (from Ref.~\cite{trans}).}
\end{center}

Figure 5 shows the significance for $H$ detection versus
$\tan\beta$ for a variety of $L/\Delta E$ possibilities: $1\rm~(fb)^{-1}/GeV$,
$0.2\rm~(fb)^{-1}/GeV$, and $0.02\rm~(fb)^{-1}/GeV$. Single $b$-tagging
with $\epsilon_b=0.5$ is assumed. Due to the sharp rise
of $n_{SD}$ with $\tanb$ a decrease of $L/\Delta E$ by a factor
of 50 only changes the lowest $\tanb$ for which discovery is possible
from $\tanb\sim 4$ to $\tanb\sim 7$.

Figure~6 provides an overall view of the possibilities for MSSM
Higgs boson detection in terms of regions in the
$\tan\beta,m_A$ plane for which $n_{SD}\geq4$ in $h=h,H,A$ searches.
This figure assumes a moderately conservative value of
$L/\Delta E=0.08\rm~(fb)^{-1}/GeV$. Further, {\it double} $b$-tagging
(with $\epsilon_b=0.6$) is required
in the $b\bar b$ mode and an additional general
efficiency factor of $0.5$ is included for both the $b\bar b$ and $t\bar t$
final states. Regions in which $h$, $H$, or $A$ detection is individually
possible are shown, as well as the additional region that is covered
by combining the $H$ and $A$ signals when they are degenerate within the final
state resolution. The figure shows that in the $b\bar b$ mode
either $H$ and $A$ detection (for $m_A\agt 140$ GeV --- as preferred
in GUT scenarios),
or $h$ and $A$ detection (for $m_A\alt 140$ GeV), will be possible provided
$\tanb> 3$--5.   Detection of $H+A$ in the $t\bar t$ mode is limited
to $m_H\sim m_A>2m_t$ and $\tanb$ values lying between 3 and about 12.

\begin{center}
\epsfxsize=3.5in\hspace{0in}\epsffile{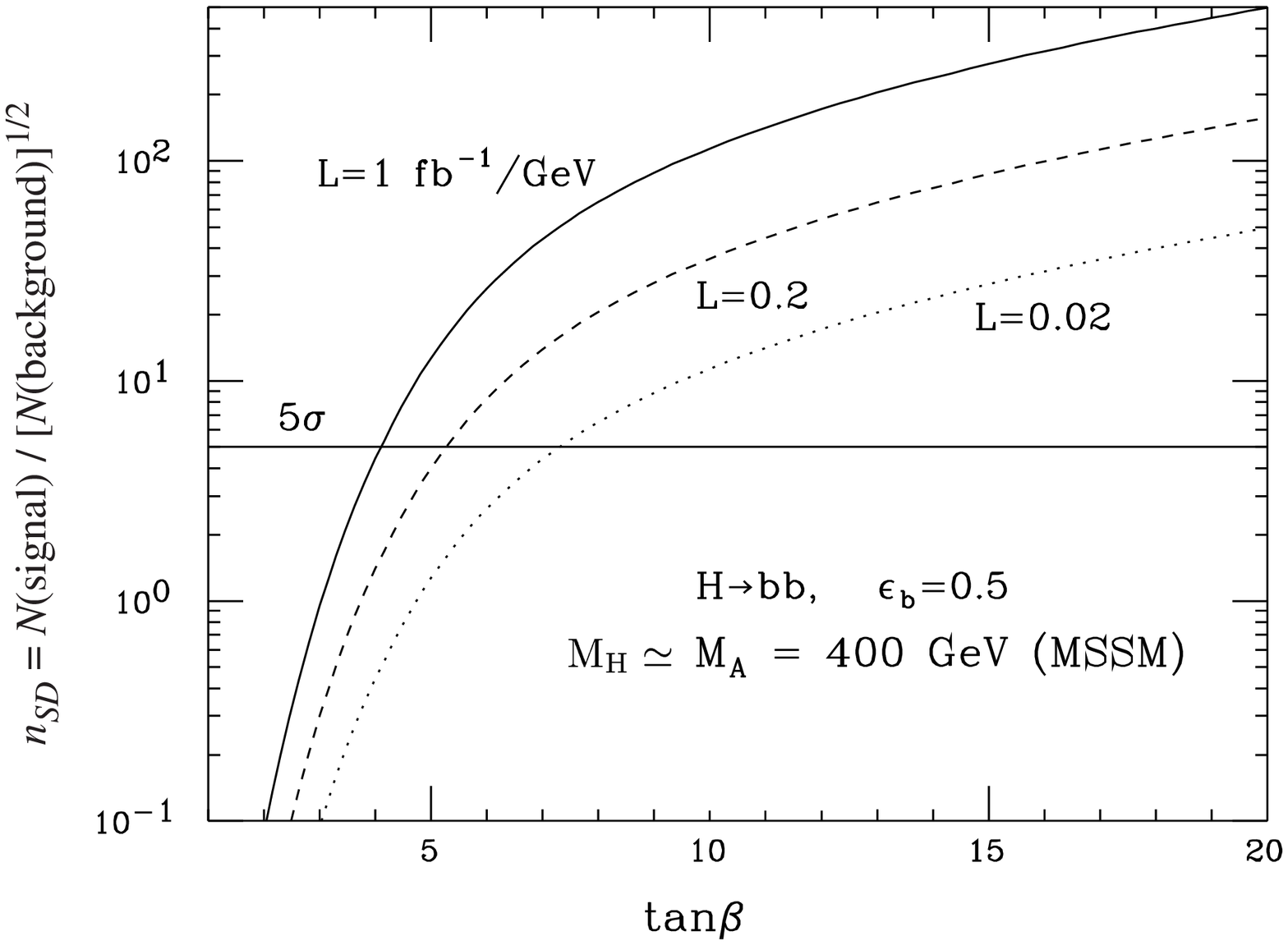}

\smallskip
\parbox{5.5in}{\footnotesize\sf Fig.~5: The statistical significance of the
Higgs $H$ signal versus $\tan \beta $. Since in SUSY GUT models one finds $H$
and $A$ to be approximately mass degenerate, the combined signal will be larger
(from Ref.~\cite{trans}).}
\end{center}

\begin{center}
\epsfxsize=3in\hspace{0in}\epsffile{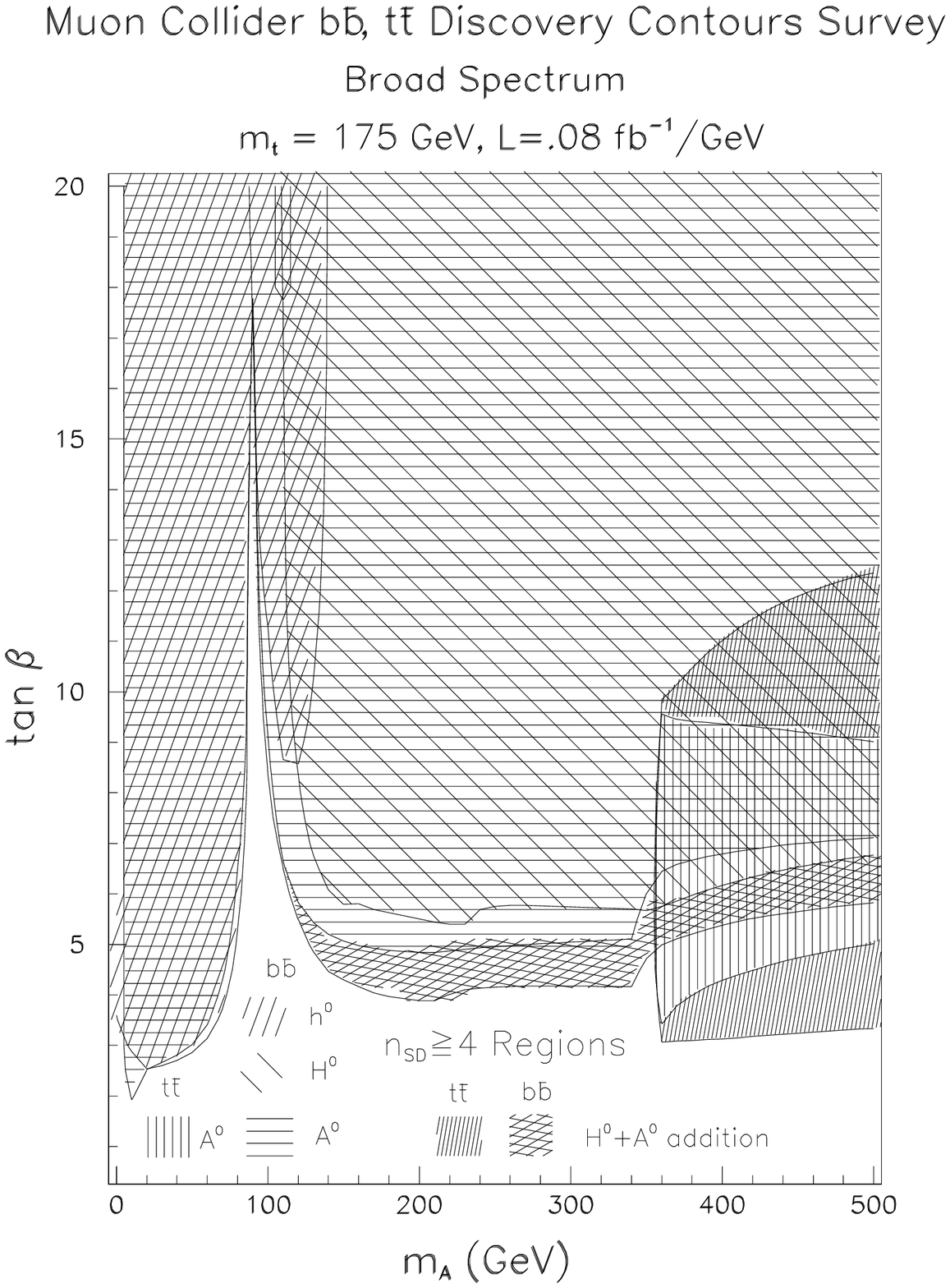}

\parbox{5.5in}{\footnotesize\sf Fig.~6: Higgs boson $H,A$ discovery regions at
a muon collider in the
modes $H,A\to b\overline {b}$ and $t\overline t$ (from Ref.~\cite{gunion}).}
\end{center}

Thus,  direct $s$-channel production allows $H,A$ discovery up
to the machine kinematical limit, so long as $\tanb$ is not small.
This is a very important extension as compared to an $\ee$ collider.
The above results assume the absence
of SUSY decay modes for the $H$ and $A$.
Once the $H,A$ mass becomes large it could happen that
SUSY decay modes become kinematically allowed. However, at large $\tanb$
the enhanced $b\bar b$ coupling guarantees that the $b\bar b$
mode branching ratio will still be large.
In practice, SUSY decay modes would only shift the
discovery regions to slightly higher $\tanb$ values.

Once a signal is identified in the search mode, the luminosity can be
concentrated over the Higgs peak, $\Delta E=2\Gamma_H\simeq 1$~GeV, to study
the resonance properties. Then the signal and background each increase by a
factor of the luminosity energy spread $\Delta E_S$ in the
broad-spectrum/scanning mode,
giving an increase in the significance of a factor of $\sqrt{\Delta E_S}$ [a
factor of 7 in the example of Eqs.~(\ref{Nsignal}--\ref{nSD})].

High polarization $P$ of both beams would be useful to suppress the background
to $s$-channel Higgs production if the luminosity reduction is less than a
factor of $\left(1+P^2\right)^2 / \left(1-P^2\right)$, which would leave the
significance of the signal unchanged~\cite{parsa}. For example, $P=0.84$ would
compensate a factor of 10 reduction in luminosity.

It could also be possible to detect the lightest SUSY Higgs boson via
$s$-channel
production if $m_h$ becomes known within a few GeV from electroweak radiative
corrections or from studies of $\mm\to Z^*\to Zh$ production, provided that the
machine
could be operated at $\sqrt s=m_h$. However, the $\mm\to h\to b\bar b$ channel
suffers a formidable background from $\mm\to\gamma^*,Z^*\to b\bar b$, so the
direct channel $h$ search might well require polarized muon beams.
Nonetheless, detection of the light $h$ in direct $s$-channel
production would be very interesting in that it would allow a determination of
the $\mm$ coupling of the $h$.

\looseness=-1
If a 0.1\% energy resolution could be achieved, another interesting application
of the FMC could be a precision determination of the top mass by measuring the
$t\bar t$ threshold cross section\cite{ttbar}.
The muon collider could present an improvement over the electron collider
from the reduced initial state radiation.

\bigskip
\begin{center}
{\large\bf\uppercase{III. Next Muon Collider} (NMC)}
\end{center}

The reduced synchrotron radiation for muons, as compared to electrons, allows
the possibility of a recirculating muon colliding beam accelerator with higher
energy than is realizable at future linear $\ee$ colliders. A design goal
under consideration is a $\sqrt s=4$~TeV $\mm$ machine with an integrated
luminosity $L\geq100\rm~(fb)^{-1}$. The construction of the higher energy $\mm$
collider could occur at the same time as the $\sqrt s=0.4$~TeV
machine~\cite{const}.

\medskip
\begin{center}
\bf A. Sparticle Studies
\end{center}

An exciting possibility is that the NMC could be a SUSY factory,
producing squark pairs, slepton pairs, chargino pairs, associated neutralinos,
associated $H+A$ Higgs, and gluinos from squark decay if kinematically allowed.
If the SUSY mass scale is $M_{\rm SUSY} \sim 1$~TeV, many sparticles could be
beyond the reach of the NLC. The LHC can produce them, but disentangling the
SUSY spectrum and measuring the sparticle masses will be a real challenge at a
hadron collider, due to the complex nature of the sparticle cascade decays and
the presence of QCD backgrounds. The measurement of the sparticle masses is
important since they are a window to GUT scale physics.

The $p$-wave suppression of squark pair production in $\ee$ or $\mm$
collisions, illustrated in Fig.~7, means that energies well above the
threshold are needed. The threshold dependence of the cross section may be
useful in sparticle mass measurements.

\vspace{-.1in}

\begin{center}
\epsfxsize=2.8in\hspace{0in}\epsffile{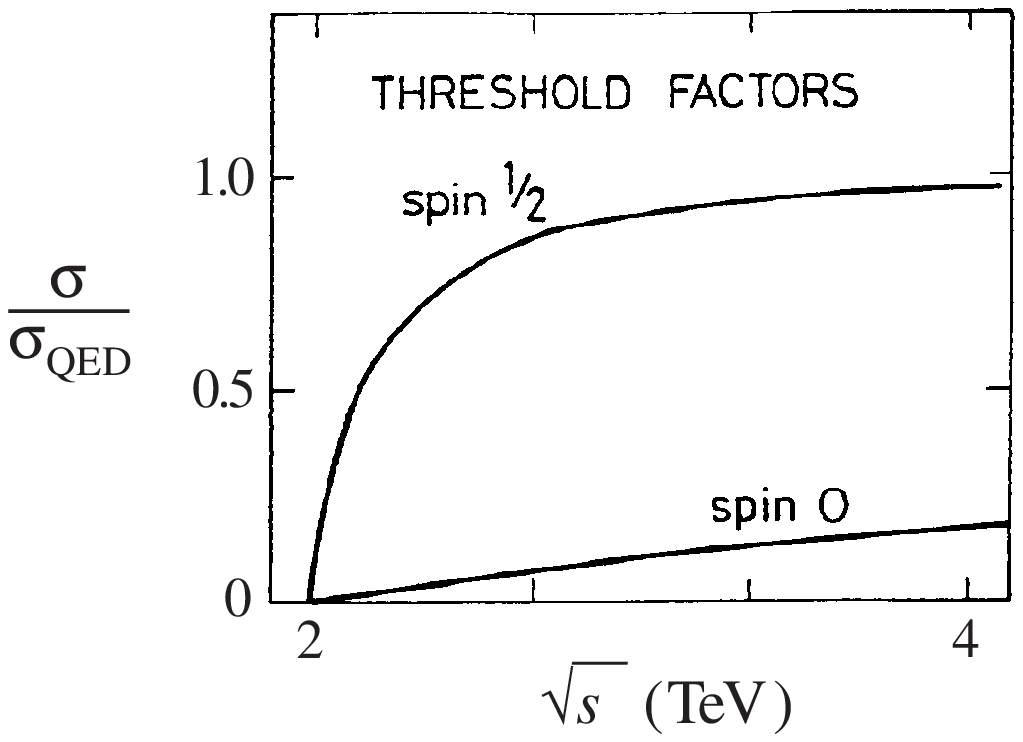}

\parbox{5.5in}{\footnotesize\sf Fig.~7: Comparison of kinematic suppression for
fermion pairs and squark pair production at $\ee$ or $\mm$ colliders.}
\end{center}

The cross sections for squarks (of one flavor in the approximation of $L,R$
degeneracy), charginos, top and three generations of singlet quarks (from an
$E_6$ GUT model, for example) are respectively
\begin{equation}  \def\arraystretch{1.25}
\begin{array}{l@{\,=\;}r@{\rm~\,fb\;\to\;}r@{\rm\ events}}
\sigma_{\tilde u_{L,R}} & 4\beta\rlap{$^3$} &  250\\
\sigma_{\tilde d_{L,R}} & 1\beta\rlap{$^3$} & 60\\
\sigma_{\chi^\pm} & 6\beta & 500\\
\sigma_t & 8 & 800\\
\sigma_{Q_{E_6}} & 6\beta & 600
\end{array}
\end{equation}
where the event rates given are for sparticle masses of 1~TeV with an
integrated luminosity of 100~fb$^{-1}$.
The production of heavy SUSY particles will give spherical events near
threshold characterized by

\begin{itemize}
\addtolength{\itemsep}{-.075in}

\item multijets

\item missing energy (associated with the LSP)

\item leptons

\end{itemize}
There should be no problem with backgrounds from SM processes.

A supergravity model with $\tan\beta=5$, universal scalar mass $m_0=1000$~GeV
and gaugino mass $m_{1/2}=150$~GeV provides an illustration of a heavy
sparticle spectrum, as follows:
\begin{equation} \def\arraystretch{1.25}
\begin{array}{c@{\qquad}r@{~}ll}
\underline{\rm sparticle} & \underline{\rm mass\vphantom{p}}\\
\tilde u & 1000 & \rm GeV\\
\tilde g & 500 \\
\tilde \ell & 1000\\
\chi_4^0,\, \chi_3^0,\, \chi_2^+ & 350\\
\chi_2^0,\,\chi_1^+ & 130\\
\chi_1^0 & 60 & & \rm (LSP)
\end{array}
\end{equation}
Consider $\tilde u \bar{\tilde u}$ production at the NMC. The dominant
cascade chain for the decays is
\begin{eqnarray}
\tilde u \bar{\tilde u} &\to& (\tilde g u)(\tilde g \bar u)\nonumber \\
\tilde g &\to& \chi_1^\pm q\bar q\\
\chi_1^\pm &\to& \chi_1^0 \ell\nu,\, \chi_1^0 q\bar q\nonumber
\end{eqnarray}
The dominant branching fractions of the  $\tilde u \bar{\tilde u}$ final state
are
\begin{equation} \def\arraystretch{1.25}
\begin{array}{r@{\rm\ jets +{} }l@{\qquad}r}
10 & \overlay p/_T & 10\%\\
8 & 1\ell + \overlay p/_T & 10\%\\
6 & 2\ell + \overlay p/_T & 2\%
\end{array}
\end{equation}
Of the two lepton events, one half will be like-sign dileptons
($\ell^+\ell^+,\, \ell^-\ell^-$). The environment of a $\mm$ collider may be
better suited than the LHC to the study of the many topologies of sparticle
events.

Turning to the SUSY Higgs sector, we simply emphasize the fact that
$Z^*\rta HA,H^+H^-$ will allow $H,A,H^{\pm}$
discovery up to $m_H\sim m_A\sim m_{H^{\pm}}$ values
somewhat below $\sqrt s/2\sim 2~\rm TeV$.  While GUT scenarios
prefer $H,A,H^{\pm}$ masses above 200 to 250 GeV, such that $HA$  and $H^+H^-$
pair production is
beyond the kinematical reach of a 400 to 500 GeV collider,
even the most extreme GUT scenarios do not yield Higgs masses beyond 2~TeV.
Thus, a 4 TeV $\mm$ collider is guaranteed to find all the SUSY Higgs bosons.

\medskip
\begin{center}
\bf B. Strong {\boldmath$WW$} Scattering
\end{center}

If Higgs bosons with
$m_H\leq{\cal O}(800\rm~GeV)$ do not exist, then the interactions of
longitudinally polarized weak bosons $W_L,\,Z_L$ became strong. This means that
new physics must be present at the TeV energy scale\cite{cg}.
The high reach in energy  of the NMC is of particular interest for study of a
strongly interacting electroweak sector (SEWS) at a
$\mm$ collider via the $WW$ fusion graphs in Fig.~8.

\begin{center}
\epsfxsize=1.8in\hspace{0in}\epsffile{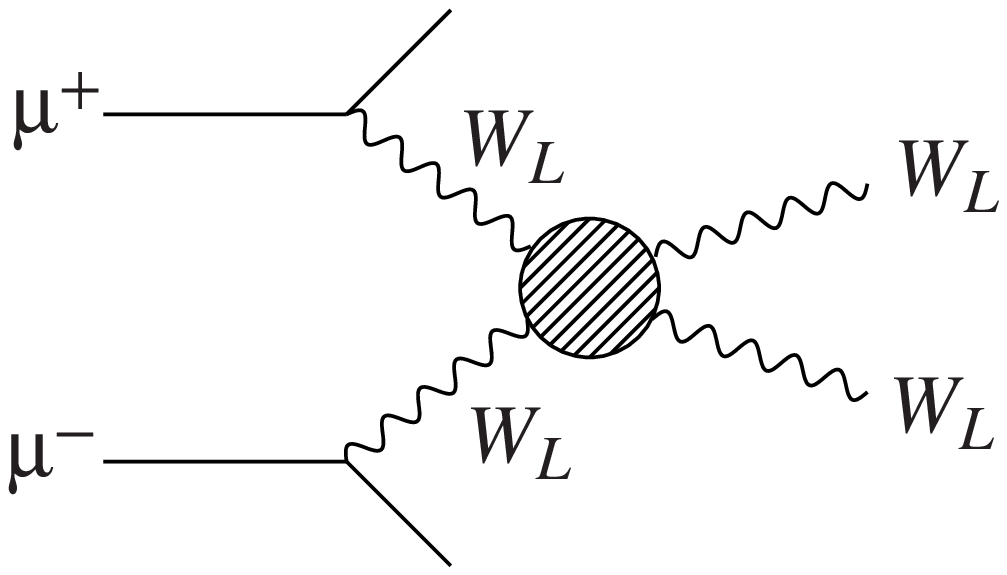}

\medskip
{\footnotesize\sf Fig.~8: Strong $W_L^+W_L^-$ scattering in $\mm$ collisions.}
\end{center}

\noindent
The SEWS signals depend on the model for $W_L^+W_L^-$ scattering. An estimate
of the size of these signals can be obtained by taking the difference of the
cross section due to a heavy Higgs boson ($m_H=1$~TeV) and that with a massless
Higgs particle
\begin{equation}
\Delta\sigma_{\rm SEWS} = \sigma(m_H=1~{\rm TeV}) - \sigma_H(m_H=0) \;.
\end{equation}
The subtraction of the $m_H=0$ result removes the contributions due to
scattering of transversely polarized $W$-bosons. Figure~9
shows the growth of
$\Delta\sigma_{\rm SEWS}$ with energy. The Table below gives results at $\sqrt
s=1.5$~TeV (for the NLC $\ee$ collider) and at $\sqrt s=4$~TeV (for the NMC).

\begin{center}
\epsfxsize=3.7in\hspace{0in}\epsffile{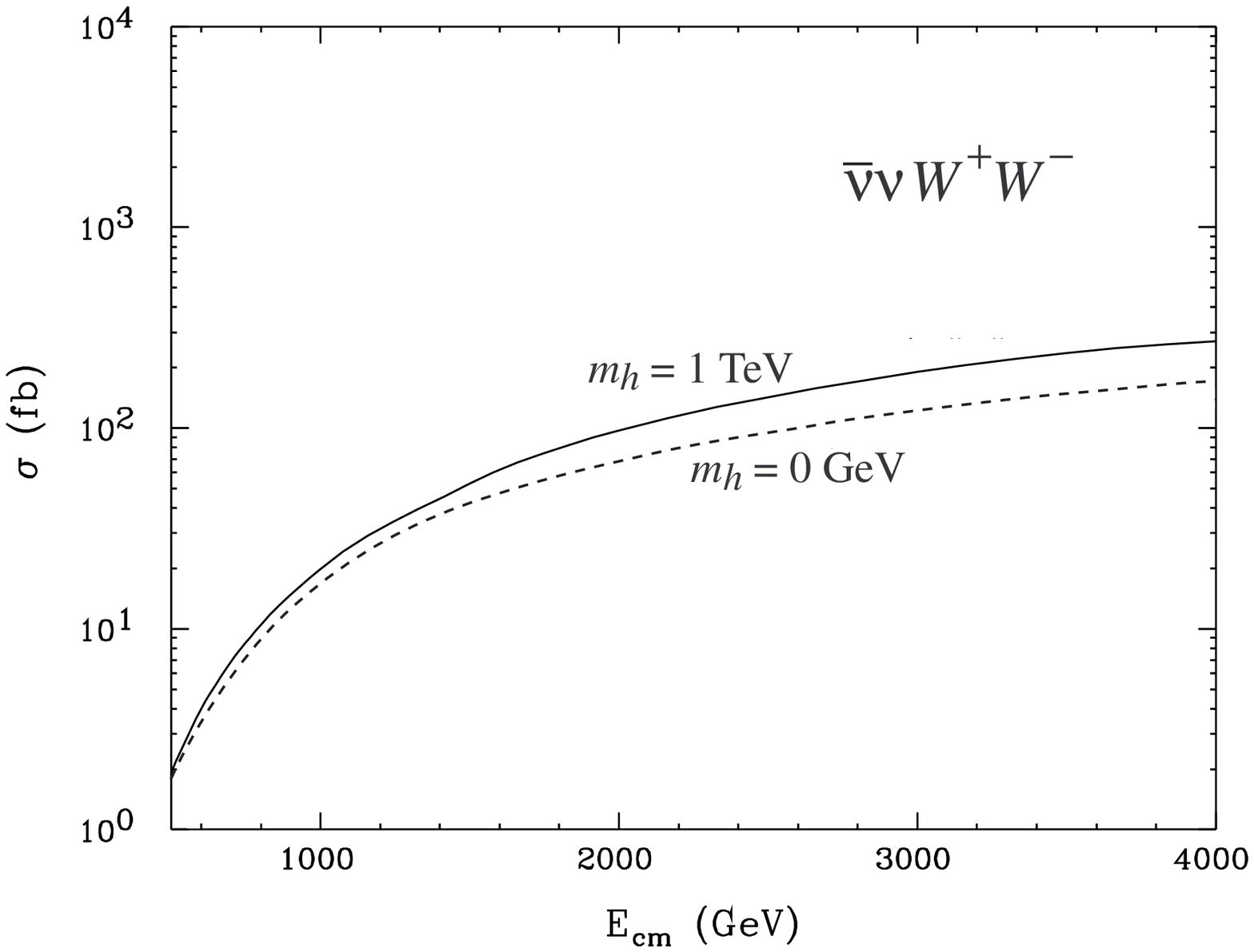}

\smallskip
{\footnotesize\sf Fig.~9: The growth of the SEWS signal with $E_{cm}$ (from
Ref.~\cite{bchp}).}
\end{center}

\medskip

\begin{center}
{\footnotesize\sf Table I: SEWS signals at future colliders}
\medskip

 \def\arraystretch{1.25}

\begin{tabular}{|c@{\quad\vline\quad}c@{\quad\vline\quad}c|}
\hline
$\sqrt s$& $\Delta\sigma(W^+W^-)$& $\Delta\sigma(ZZ)$\\ \hline
1.5 TeV& 8 fb& 6 fb\\
4 TeV& 80 fb& 50 fb\\ \hline
\end{tabular}
\end{center}

The energy reach is thereby a critical consideration in the study of SEWS.
Additionally, the backgrounds
from the photon exchange process of Fig.~10 will
be a factor of 3 less at a $\mm$ machine compared to an $\ee$
machine\cite{bchp}. The
probability for $\gamma$ radiation from a charged lepton
\begin{equation}
P_{\gamma/\ell}(x) = {\alpha\over\pi} {1+(1-x)^2\over x} \ln(E_\ell/m_\ell)
\end{equation}
is logarithmically dependent on the charged lepton mass. Figure~11
shows SM cross sections with $m_H=0$ versus CM energy.
The background to SEWS from $\ee\to \ee W^+W^-$
can be rejected by requiring\cite{hkm}
\begin{equation}
\mbox{no $e^\pm$ with $E_\ell>50$ GeV and
$|\cos\theta_\ell|<|\cos(0.15$ rad)}|\;.
\end{equation}
Some similar rejection would be necessary to suppress the $\mm\to\mm W^+W^-$
background at the NMC.

\begin{center}
\epsfxsize=1.8in\hspace{0in}\epsffile{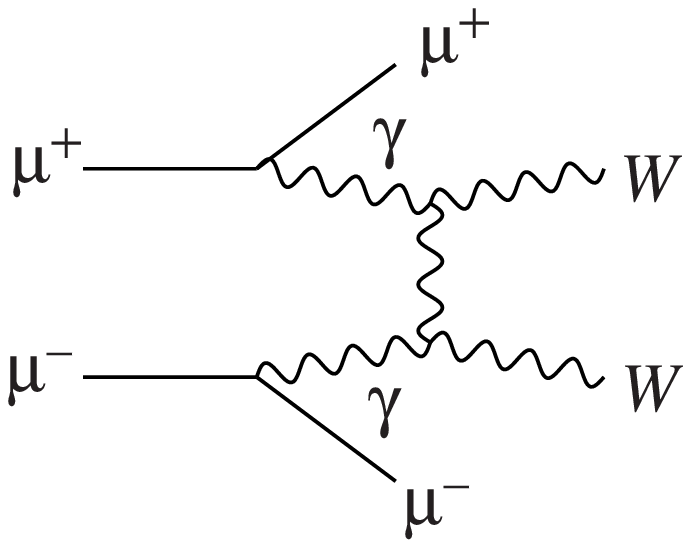}

{\footnotesize\sf Fig.~10: The background to the SEWS signal
from the photon exchange process.}
\end{center}

\smallskip

\begin{center}
\epsfxsize=4in\hspace{0in}\epsffile{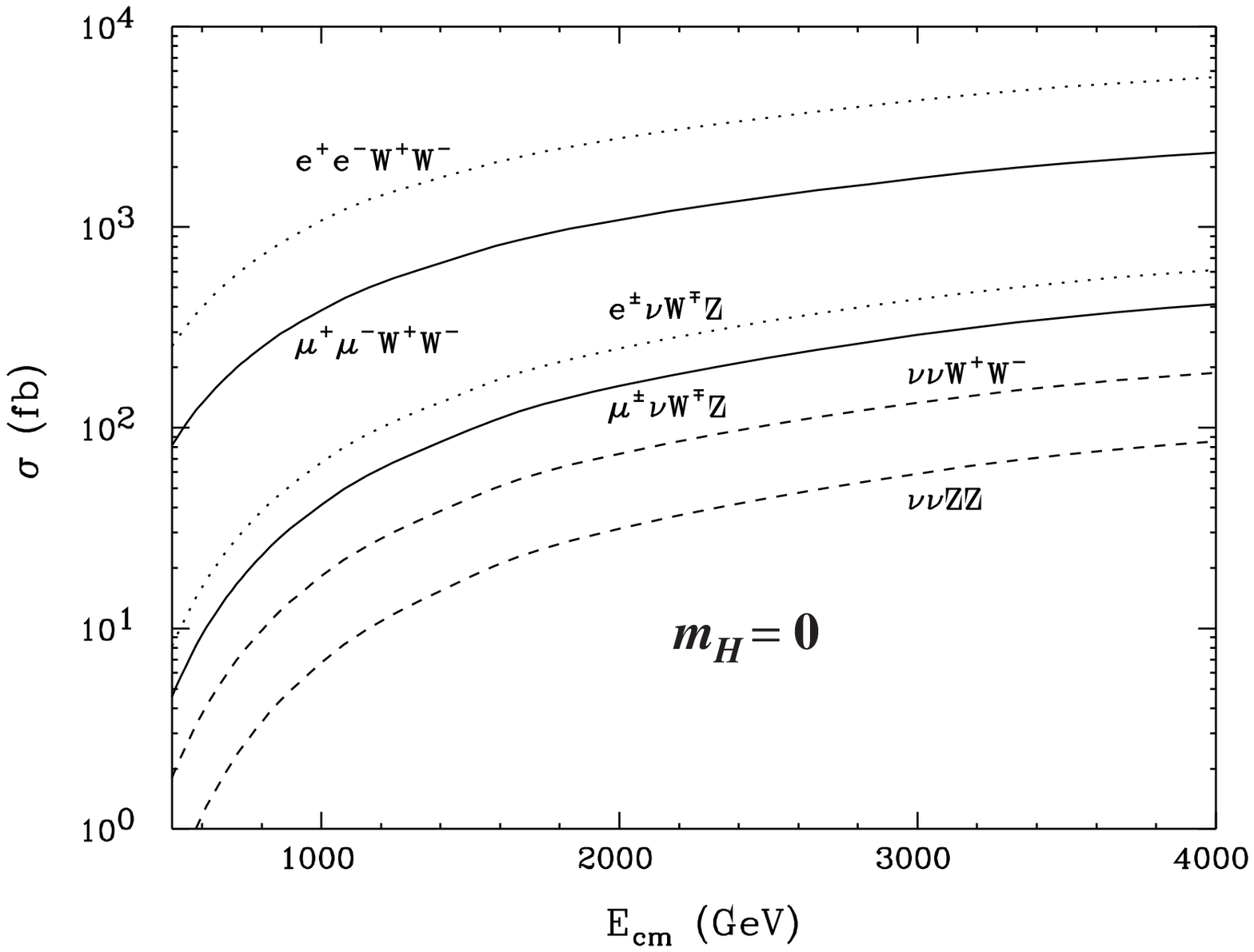}

{\footnotesize\sf Fig.~11: The cross sections for Standard Model processes at a
muon
collider (from Ref.~\cite{bchp}).}
\end{center}

The various SM cross sections grow with energy, for
example,
\begin{equation}
\sigma(\mm\to\mm W^+W^-) = 2000\rm\ fb
\end{equation}
at $\sqrt s=4$~TeV. These cross sections would be higher if the $W$ is
composite. Thus precision studies of $WW$ self couplings should be possible at
the NMC.

There are many other new possibilities of interest for the NMC that we have not
yet addressed, including

\begin{itemize}

\item extra neutral gauge bosons (the NMC could be a $Z'$ factory, with its
decays giving Higgses and $W^+W^-$ along with particle and sparticle pairs)

\item right-handed weak bosons (the present limit on the right-handed weak
boson of left-right symmetric models is $M_{W_R}\agt 1.5$~TeV)

\item vector-like quarks and leptons (present in $E_6$ models)

\item horizontal gauge bosons $X$ (whose presence may be detected as an
interference between $t$-channel $X$ exchange and $s$-channel $\gamma,Z$
exchanges; present limits are $M_X\agt1$~TeV)

\item leptoquarks

\end{itemize}

\noindent
The list goes on with other exotica.

\medskip
\begin{center}
\large\bf IV. CONCLUSION
\end{center}

In conclusion, $\mm$ colliders seem to offer unparalleled new opportunities at
both the low $(\sqrt s\simeq400$~GeV) and high $(\sqrt s\simeq 4$~TeV) energy
frontiers. The primary advantages of such a collider are:

\begin{itemize}

\item to discover and study Higgs bosons that are not coupled to $ZZ$ or $WW$
(e.g.\ the heavy SUSY Higgs bosons $H,A$) by employing either direct
$s$-channel resonance production (at a low energy machine) or $Z^*\rta HA$
(at a high energy machine);

\item to measure the masses and properties of heavy SUSY particles in the
improved background environment of a lepton collider;

\item to study a strongly interacting electroweak sector with higher signal
rates at higher energies.

\end{itemize}

Along with accelerator development, much work remains to be done on the physics
for such machines and a continuing investigation is underway\cite{bbgh}.

\bigskip\goodbreak
\begin{center}
{\large\bf\uppercase{Acknowledgments}}
\end{center}

This work was supported in part by the U.S.~Department of Energy under Grants
No.~DE-FG02-95ER40896 and No.~DE-FG03-91ER40674. Further support was provided
by the University of Wisconsin Research
Committee, with funds granted by the Wisconsin Alumni Research Foundation,
and by the Davis Institute for High Energy Physics.

\end{document}